\documentclass[twocolumn,showpacs,aps,prl,superscriptaddress]{revtex4}

\usepackage{graphicx}
\usepackage{dcolumn}
\usepackage{amsmath}
\usepackage{epsfig}
\usepackage{bm}

\def\ra{\rightarrow}
\def\sigpll{\Sigma^+ \rightarrow p l^+ l^-}

\def\sigpee{\Sigma^{+}_{p e e}}
\def\spmumu{\Sigma^{+}_{p \mu \mu}}
\def\sigp_ppi0{\Sigma^+ \rightarrow p \pi^0}
\def\sppi0{\Sigma^{+}_{p \pi^0}}
\def\speeg{\Sigma^{+}_{p e e \gamma}}

\def\kpieeg{K^{+}_{\pi e e \gamma}}
\def\sppmumu{\Sigma^{+}_{p P \mu \mu}}

\begin{document}

\title{Evidence for the  Decay {\boldmath$\Sigma^+ \rightarrow p \mu^{+} \mu^{-}$}}

\affiliation{Institute of Physics, Academia Sinica, Taipei 11529, Taiwan,
             Republic of China}
\affiliation{University of California, Berkeley, California 94720, USA}
\affiliation{Fermi National Accelerator Laboratory, Batavia, Illinois 60510, USA}
\affiliation{Universidad de Guanajuato, 37000 Le\'{o}n, Mexico}
\affiliation{Illinois Institute of Technology, Chicago, Illinois 60616, USA}
\affiliation{Universit\'{e} de Lausanne, CH-1015  Lausanne, Switzerland}
\affiliation{Lawrence Berkeley National Laboratory, Berkeley, California 94720, USA}
\affiliation{University of Michigan, Ann Arbor, Michigan 48109, USA}
\affiliation{University of South Alabama, Mobile, Alabama 36688, USA}
\affiliation{University of Virginia, Charlottesville, Virginia 22904, USA}

\author{H.K.~Park}
\affiliation{University of Michigan, Ann Arbor, Michigan 48109, USA}
\author{R. A. Burnstein}
 \affiliation{Illinois Institute of Technology, Chicago, Illinois 60616, USA}
\author{A. Chakravorty}
 \affiliation{Illinois Institute of Technology, Chicago, Illinois 60616, USA}
\author{Y. C. Chen}
 \affiliation{Institute of Physics, Academia Sinica, Taipei 11529, Taiwan,
              Republic of China}
\author{W. S. Choong}
 \affiliation{University of California, Berkeley, California 94720, USA}
 \affiliation{Lawrence Berkeley National Laboratory, Berkeley, California 94720, USA}
\author{K. Clark}
 \affiliation{University of South Alabama, Mobile, Alabama 36688, USA}
\author{E. C. Dukes}
 \affiliation{University of Virginia, Charlottesville, Virginia 22904, USA}
\author{C. Durandet}
 \affiliation{University of Virginia, Charlottesville, Virginia 22904, USA}
\author{J. Felix}
 \affiliation{Universidad de Guanajuato, 37000 Le\'{o}n, Mexico}
\author{Y. Fu}
 \affiliation{Lawrence Berkeley National Laboratory, Berkeley, California 94720, USA}
\author{G. Gidal}
 \affiliation{Lawrence Berkeley National Laboratory, Berkeley, California 94720, USA}
\author{H. R. Gustafson}
 \affiliation{University of Michigan, Ann Arbor, Michigan 48109, USA}
\author{T. Holmstrom}
 \affiliation{University of Virginia, Charlottesville, Virginia 22904, USA}
\author{M. Huang}
 \affiliation{University of Virginia, Charlottesville, Virginia 22904, USA}
\author{C. James}
 \affiliation{Fermi National Accelerator Laboratory, Batavia, Illinois 60510, USA}
\author{C. M. Jenkins}
 \affiliation{University of South Alabama, Mobile, Alabama 36688, USA}
\author{T. Jones}
 \affiliation{Lawrence Berkeley National Laboratory, Berkeley, California 94720, USA}
\author{D. M. Kaplan}
 \affiliation{Illinois Institute of Technology, Chicago, Illinois 60616, USA}
\author{L. M. Lederman}
\affiliation{Illinois Institute of Technology, Chicago, Illinois 60616, USA}
\author{N. Leros}
\affiliation{Universit\'{e} de Lausanne, CH-1015 Lausanne, Switzerland}
\author{M. J. Longo}
 \email[Corresponding author. \\ 
        Email: ]{mlongo@umich.edu}
 \affiliation{University of Michigan, Ann Arbor, Michigan 48109, USA}
\author{F. Lopez}
\affiliation{University of Michigan, Ann Arbor, Michigan 48109, USA}
\author{L. C. Lu}
 \affiliation{University of Virginia, Charlottesville, Virginia 22904, USA}
\author{W. Luebke}
 \affiliation{Illinois Institute of Technology, Chicago, Illinois 60616, USA}
\author{K.B. Luk}
 \affiliation{University of California, Berkeley, California 94720, USA}
 \affiliation{Lawrence Berkeley National Laboratory, Berkeley, California 94720, USA}
\author{K. S. Nelson}
 \affiliation{University of Virginia, Charlottesville, Virginia 22904, USA}
\author{J.-P. Perroud}
 \affiliation{Universit\'{e} de Lausanne, CH-1015 Lausanne, Switzerland}
\author{D. Rajaram}
 \affiliation{Illinois Institute of Technology, Chicago, Illinois 60616, USA}
\author{H. A. Rubin}
 \affiliation{Illinois Institute of Technology, Chicago, Illinois 60616, USA}
\author{J. Volk}
 \affiliation{Fermi National Accelerator Laboratory, Batavia, Illinois 60510, USA}
\author{C. G. White}
 \affiliation{Illinois Institute of Technology, Chicago, Illinois 60616, USA}
\author{S. L. White}
 \affiliation{Illinois Institute of Technology, Chicago, Illinois 60616, USA}
\author{P. Zyla}
 \affiliation{Lawrence Berkeley National Laboratory, Berkeley, California 94720, USA}

\collaboration{HyperCP Collaboration}
\noaffiliation

\date{\today}

\begin{abstract}
\vspace{0.1in}
We report the first evidence for the decay $\Sigma^+ \ra p \mu^+ \mu^-$ 
from data taken by the HyperCP (E871) experiment at Fermilab. 
Based on three observed events, the
branching ratio is $\mathcal{B}(\Sigma^+ \ra p \mu^+ \mu^-) = 
[8.6^{+6.6}_{-5.4}({\rm stat}) \pm 5.5({\rm syst})] \times 10^{-8}$. 
The narrow range of dimuon masses may indicate that the decay
proceeds via a neutral intermediate state, $\Sigma^+ \ra p P^0, P^0 \ra \mu^+
\mu^-$ with a $P^0$ mass of $214.3 \pm 0.5$ MeV/$c^2$ and branching ratio
$\mathcal{B}(\Sigma^+ \ra p P^0, P^0 \ra \mu^+ \mu^-) = [3.1^{+2.4}_{-1.9}({\rm stat}) \pm
1.5({\rm syst})]  \times 10^{-8}$.
\end{abstract}


\maketitle

\begin{figure}[htb]
\centerline{\psfig{figure=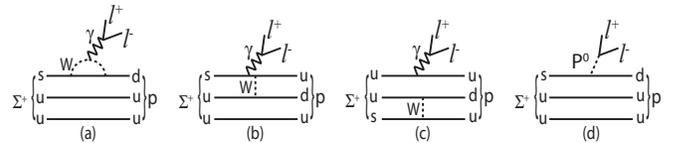,width=3.4in}}
\caption{Feynman diagrams for $\Sigma^{+}_{pll}$ decays in the SM   
(a)--(c) and via new physics (d). The SM processes are referred to 
as FCNC (a) and internal conversion (b)--(c).}
\label{fig:feyn_sig}
\end{figure}

In the standard model (SM), the decay $\Sigma^+ \ra p l^+
l^-~(\Sigma^{+}_{pll}, l=e, \mu)$ can be described as proceeding through a
flavor-changing neutral-current (FCNC) interaction and by internal conversion,
as shown in Fig.~\ref{fig:feyn_sig}(a)--(c). 
Bergstr\"om {\sl et al.}~\cite{bergstr} argue that in the SM
the FCNC contribution for the decay $\Sigma^{+}_{pll}$ is not dominant.
The decay $\Sigma^{+}_{pll}$ is of
interest since it also allows a direct search for a new scalar or
vector particle, which could contribute an $s \ra d$ transition at the tree
level~\cite{gor} (Fig.~\ref{fig:feyn_sig}(d)). 

Current literature reports only 
an upper limit $\mathcal{B}(\Sigma^+ \ra p e^+ e^-) < 7
\times 10^{-6}$~\cite{ang}. The decay rate for the process $\Sigma^+ \ra p l^+
l^-$ was studied in~\cite{bergstr,corrigan} 
in the context of the SM.
Using the measured partial decay width and 
the $\alpha_{\gamma}$ parameter 
for $\Sigma^+ \ra p \gamma$ decay, Bergstr\"om {\sl et al.}~\cite{bergstr} 
find: $\Gamma(\Sigma^+ \ra p e^+ e^-)/\Gamma(\Sigma^+ \ra p \gamma) \ge 7.2
\times 10^{-3}$  and $1/120 \gtrsim \Gamma(\Sigma^+ \ra p \mu^+ \mu^-) /
\Gamma(\Sigma^+ \ra p e^+ e^-)  \gtrsim 1/1210$.
A large violation of these limits would signal new physics.

In this Letter we report  the first evidence for the $\Sigma^+ \ra p
\mu^+ \mu^-$ decay, a measurement of the branching ratio for this decay, 
and possible evidence for a new state $P^0$ observed via 
$\Sigma^+ \ra p P^0$, $P^0 \ra \mu^+ \mu^-$.

\begin{figure}[htb]
\centerline{\psfig{figure=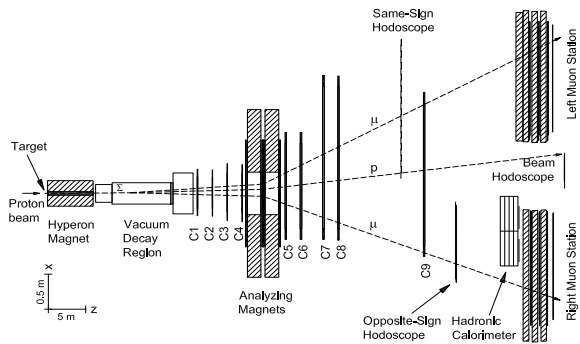,height=45mm,width=3.0in}}
\caption{Plan view of the HyperCP spectrometer.}
\label{fig:spect}
\end{figure}

The HyperCP experiment was located in the Meson Center beamline at Fermilab.
The spectrometer, shown in Fig.~\ref{fig:spect}, is described in detail 
elsewhere~\cite{burnstein}.
Charged secondary beams, with mean momenta of about 160 GeV/$c$,
were produced by 800 GeV/$c$ protons incident on copper targets
and momentum selected by a 
curved collimator situated in a dipole magnet (hyperon magnet).  
The sign of the charged secondary beam was periodically changed 
by reversing the field of the hyperon magnet.
We analyzed $2.14 \times 10^9$ triggers from the positive-secondary-beam 
data set and 
$0.37 \times 10^9$
from the negative.

The signature of the $\spmumu$ decay is two unlike-sign muon tracks and a
proton track originating from a common vertex. 
The transverse-momentum kick of the
analyzing magnets was such that muons from this decay were always deflected to
opposite sides and outside of the intense secondary beam at 
the rear of the spectrometer. 
Hence the signal trigger used to select candidate events 
required hits in the same-sign and 
opposite-sign hodoscopes (Left-Right trigger),
in coincidence with hits in the vertical
and horizontal hodoscopes in each of the muon stations situated on
either side of the secondary beam.

The basic selection cuts discussed below were applied to the $\spmumu$ event
candidates. 
Each event was required to have a track in each muon station
and a higher-momentum track (the proton candidate) with the same 
charge sign as the secondary beam.
The total momentum of the three tracks had to be within the range
120 to 240 GeV/$c$. The $\Sigma^+$ trajectory had to extrapolate to within 3.5
mm ($\approx 3\,\sigma$) of the center of the target, 
where the nominal target
position was determined using well-reconstructed 
$K^+ \ra \pi^+ \pi^+ \pi^-$ decays and the extrapolation resolution
was determined from a Monte Carlo (MC) simulation of $\spmumu$ decays. 
The decay vertex of the three tracks was
calculated by the method of distance of closest approach (DCA), and the
$z$-coordinate of the vertex $z_{v}$ was required to be
more than 68 cm downstream of the entrance of the vacuum decay region and
more than 32 cm upstream of its exit.
The average distance between pairs of tracks in the $x$--$y$ plane 
at $z_v$  was required to be less than 0.25 mm. 
The hits in the multiwire proportional chambers upstream of the 
analyzing magnets were refit with a constraint that they share 
a common vertex. The resulting $\chi^2/ndf$ was
required to be less than 1.5.

\begin{figure}[htb]
\centerline{\psfig{figure=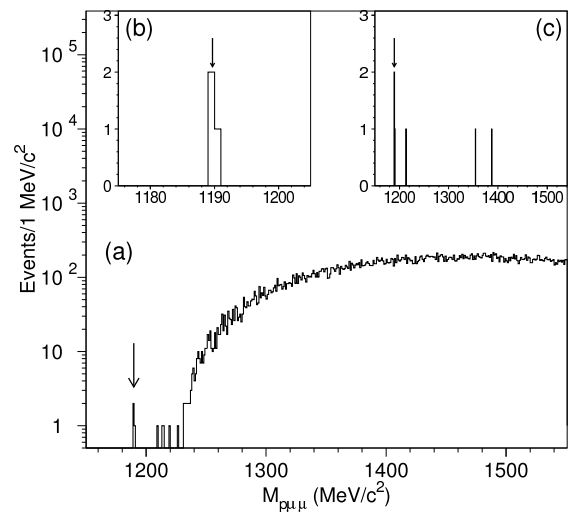,width=2.9in}}
\caption{$M_{p\mu\mu}$ distribution
for the positive-secondary-beam data,
(a) after the standard cuts,
(b) within $\pm 15$ MeV/$c^2$ of the  $\Sigma^+$ mass,
and (c) after the additional cuts.
The arrow represents the $\Sigma^+$ mass.}
\label{fig:pmumu}
\end{figure}

After imposition of the above cuts, 
three candidate $\spmumu$ events were observed
in the positive-secondary-beam data
with masses ($M_{p\mu\mu}$) within $1\,\sigma$ (1 MeV/$c^2$) of 
the $\Sigma^+$ mass, as
shown in Fig.~\ref{fig:pmumu}(a) and (b). No other events were found within
$\approx 20$ MeV/$c^2$ of the $\Sigma^+$ mass.

The signal events were verified by
applying two additional cuts that removed almost all of the higher-mass
background events without affecting the three signal events.
The first cut required that the ratio of the proton momentum to 
the summed three-track momentum be larger than 0.68, as
MC simulations showed that this cut preserved 100\% of the signal 
while removing most of the $K^+$ decay background. 
The second cut removed events whose mass was within $\pm 10$ MeV/$c^2$
($3\,\sigma$) of the $K^+$ mass when calculated using the
$\pi^+ \mu^+ \mu^-$ decay hypothesis to reject
$K^+ \ra \pi^+ \mu^+ \mu^-$ decays.
The resulting $p\mu^+\mu^-$ invariant mass distribution is 
shown in Fig~\ref{fig:pmumu}(c).

Possible background sources were extensively studied, using both MC
and data.
Typically 100 times as many MC events as would be expected 
for each potential background source were generated.

Other positively-charged hyperon decays, 
$\overline{\Xi}$$^+ \ra \overline{\Lambda} \pi^+ \ra \overline{p}\pi^+\pi^+$ 
and
$\overline{\Omega}$$^+ \ra \overline{\Lambda} K^+ \ra \overline{p}\pi^+ K^+$, 
were not significant backgrounds
as decays in flight of the daughter $\pi^+$ or $K^+$ would 
produce muons of only one polarity, and
the probability of misidentifying the sign of the charge
of any of the decay daughters was negligibly small.
Contributions from charged kaon decays such as $K^+ \ra \pi^+ \pi^+ \pi^-$,
$K^+ \ra \pi^+ \pi^- \mu^+ \nu_{\mu}$, $K^+ \ra \pi^+ \mu^+ \mu^-$, and
$K^+ \ra \mu^+ \mu^+ \mu^- \nu_{\mu}$ were estimated using MC simulations
allowing in-flight decay of pions, and found to be negligible.
Muon pair production by photon conversion in material
inside the vacuum decay region was studied using MC simulation
for $K^+ \ra \pi^+ \pi^0, \pi^0 \ra \gamma \gamma$, $K^+ \ra \pi^+ \gamma
\gamma$, $\Sigma^+ \ra p \gamma$, and $\Sigma^+ \ra p  \pi^0, \pi^0 \ra \gamma
\gamma$ decays.  Such sources of background were also found to be
negligible.
 
In addition, we used real data to investigate
possible backgrounds otherwise missed, including the unlike-sign dimuon sample
from the negative-secondary-beam data, as well as a sample of events 
(single-muon sample) for which
only one muon track was required in either the left or right muon
station.  For the single-muon sample
both the positive- and negative-secondary-beam data were included,
and the non-muon track was required to be 
within the fiducial volume of the appropriate muon station. 
For the positive-secondary-beam data, the single-muon 
sample was one order of magnitude larger than 
the unlike-sign dimuon sample.
These background studies
showed that after cuts none of these sources contributed in 
the $p\mu^+\mu^-$ invariant-mass region below 1200 MeV/$c^2$.
Finally, we relaxed each cut  to
increase the background level shown in Fig.~\ref{fig:pmumu}(c) by an order of
magnitude. However, there still were no background events within 8 MeV/$c^2$
of the $\Sigma^+$ mass.

\begin{figure}[h]
\centerline{\psfig{figure=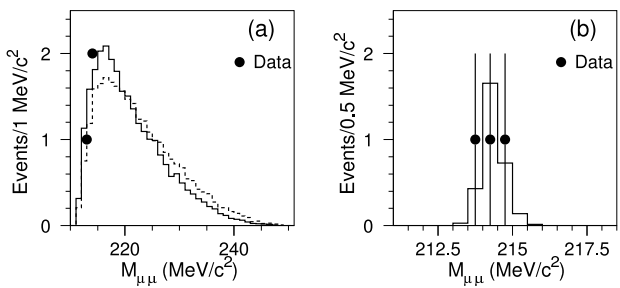,width=3.2in}}
\caption{Real (points) and MC (histogram) dimuon mass distributions for  
    (a) $\spmumu$ MC events (arbitrary normalization) 
with a form-factor decay (solid histogram)
and uniform phase-space decay (dashed histogram) model,
and (b) $\sppmumu$ MC events normalized to match the data.}
\label{fig:mumu}
\end{figure}

Figure~\ref{fig:mumu}(a) compares the dimuon mass distribution
of the three signal candidates with that 
expected in the SM with the form-factors described below. 
The reconstructed dimuon masses for the three candidates, 214.7,
214.3, and 213.7 MeV/$c^2$, all lie within the expected dimuon mass
resolution of $\approx 0.5$ MeV/$c^2$. 
The dimuon mass distribution for $\spmumu$ decays is expected to be broad 
unless the form factor  has a pole in the kinematically allowed range of 
dimuon mass.

The expected SM distribution 
was used to estimate the probability that the 
dimuon masses of the three signal candidates be within 1 MeV/$c^2$ 
of each other anywhere within the kinematically allowed range. 
The probability is 0.8\% for the form-factor decay model and
0.7\% for the uniform phase-space decay model. 
The unexpectedly narrow dimuon mass distribution suggests a two-body 
decay, $\Sigma^+ \ra p P^0, P^0 \ra \mu^+ \mu^-$ ($\sppmumu$), 
where $P^0$ is an unknown particle with mass $214.3 \pm 0.5$ MeV/$c^2$. 
The dimuon mass distribution for the three signal candidates is compared with 
MC $\sppmumu$ decays in Fig.~\ref{fig:mumu}(b), 
and good agreement is found. 
Distributions of hit positions and momenta of the proton, $\mu^+$, and $\mu^-$
of the three candidate events were compared with MC distributions,
and were found to be consistent with both decay hypotheses. 

To extract the $\spmumu$ branching ratio, the 
$\Sigma^+ \ra p \pi^0, \pi^0 \ra e^+ e^- \gamma$ ($\speeg$) decay was used
as the normalization mode, where the $\gamma$ was not detected. (HyperCP
had no $\gamma$ detectors.) 
The trigger for the $\speeg$ events was the Left-Right trigger
prescaled by 100. 
The proton and two unlike-sign electrons were required to come
from a single vertex, as were the three tracks of the signal mode.

The proton was selected to be the positively-charged track with the 
greatest momentum, and the event was discarded if the proton
candidate did not have at least 66\% of the total 
three-track momentum, as determined by a MC simulation of $\speeg$ decays. 
The reconstructed
mass for the ${3\pi}$ hypothesis was required to be outside $\pm 10$
MeV/$c^2$ of the $K^+$ mass. 
The cuts on $\chi^2/ndf$, DCA, and the
total momentum were the same as for the signal mode. However, the decay vertex
had to be more than 168 cm downstream of 
the entrance of the vacuum decay region
and more than 32 cm upstream of its exit. 
Since the $\gamma$ momentum was not measured, the $x$ and $y$
positions of the $\Sigma^+$ trajectory at the target 
were determined using only the three charged tracks, 
and those positions had to be consistent with that expected from a MC
simulation of $\speeg$ decays.

\begin{figure}[htb]
\centerline{\psfig{figure=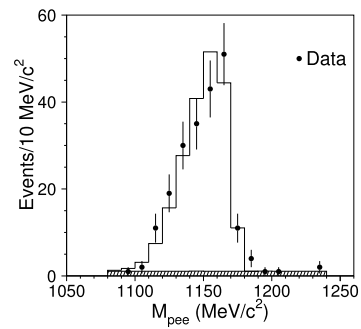,width=1.80in}}
\caption{The reconstructed $p e^+ e^-$ mass distribution
for the normalization mode after all cuts.  
The histogram is the sum of MC samples of $\speeg$, $\kpieeg$ decays
and a uniform background, where the relative amounts of each were
determined by a fit, and the number of MC events was normalized to
match the number of data events. 
The hatched area shows the main background source (uniform background).
}
\label{fig:pee}
\end{figure}

To significantly reduce contamination
from photon-conversion events, the dielectron mass was required to be between
50 and 100 MeV/$c^2$. After application of the
above selection criteria, a total of
211 events remained, as shown in Fig.~\ref{fig:pee}. We performed a binned
maximum-likelihood fit for the mass distributions for data 
and three MC samples:
$\speeg$ decays, $K^+ \ra \pi^+ \pi^0$, $\pi^0 \ra e^+ e^- \gamma$ ($\kpieeg$) 
decays, and uniform background. 
From the fit, the number of observed $\speeg$ decays was 
$N^{obs}_{nor}=189.7 \pm 27.4$ events, where the uncertainty is statistical. 
To extract the total number of normalization events, values of 
$(51.57{\pm}0.30)$\% and $(1.198{\pm}0.032)$\% were used respectively
for the $\Sigma^+ \ra p \pi^0$ and $\pi^0 \ra e^+ e^- \gamma$ branching
ratios \cite{pdg}.

The kinematic parameters for $\Sigma^+$ production at the target were tuned
to match the data and MC $\speeg$ momentum distributions. 
The MC $\speeg$ decays were generated using the decay model 
in Ref.\ \cite{mikaelian} for $\pi^0 \ra e^+ e^- \gamma$ ($\pi^0_{ee\gamma}$)
decays, and the
$\pi^0$ electromagnetic form-factor parameter $a = 0.032 \pm 0.004$ was
taken from Ref.\ \cite{pdg}.
After tuning of the parameters, comparisons of the distributions of
the MC events with the data for $\speeg$ decays, the decay vertex positions,
momentum spectra, reconstructed mass, hit positions of each charged
particle, etc., showed good agreement.

In the simulation of the $\spmumu$ decays, we used 
the form-factor model of Bergstr\"{o}m {\em et al}.\ \cite{bergstr},
although we found little difference between results using it and 
\begin{table}[htpb]
\centering
\caption{Acceptances ($A$) and
efficiencies ($\epsilon$) for the signal and normalization modes.}
\label{tab:accep}
\begin{ruledtabular}
\begin{tabular}{lrr}
\multicolumn{1}{c}{Mode}          
&  \multicolumn{1}{c}{$A$ ($\%$)}
&  \multicolumn{1}{c}{$\epsilon$ ($\%$)} \\ \hline
$\spmumu$ decays           & 0.259 & 71.2         \\
$\sppmumu$ decays          & 0.731 & 69.4         \\
Normalization $\speeg$     & 0.255 & 5.6         \\ 
\end{tabular}
\end{ruledtabular}
\end{table}
a uniform phase-space decay model.
The form-factor model uses the SM processes (FCNC and internal conversion)
shown in Fig.~\ref{fig:feyn_sig}(a)--(c).
In the model $\mathcal{B}(\sigpll)$ depends on three
parameters: $(b_2/b_1)$, $(c_1/b_1)$ and $(c_2/b_1)$.
Using the Particle Data Group values of the partial decay width and 
the $\alpha_{\gamma}$ decay 
parameter of the $\Sigma^+ \ra p \gamma$ decay, we determined 
$(b_2/b_1)=-0.46 \pm 0.07$. 
The other two parameters were chosen to make the $\sigpee$ branching
ratio as small as possible so that it is consistent with the
experimental upper limit, $< 7 \times 10^{-6}$~\cite{ang,pdg}.
This gives $(c_1/b_1)=0.7$ and $(c_2/b_1)=-13.0$, and 
a branching ratio  
$\mathcal{B}(\Sigma^+ \ra p e^+ e^-)=8.9 \times 10^{-6}$.

In the simulation of the $\sppmumu$ decay, we assumed that the $P^0$ mass was
214.3 MeV/$c^2$ with negligible decay width, and that the $P^0$ decayed
immediately to the dimuon pair. The matrix element for 
the $P^0 \ra \mu^+ \mu^-$ decay was taken to be uniform.

MC simulations were used to estimate the geometric acceptances and the
event-selection efficiencies for the signal and normalization modes.
Table~\ref{tab:accep} shows a summary of the results. 
The relative trigger
efficiency for the signal mode with respect to the normalization mode and the
unlike-sign dimuon detection efficiency were determined from the 
full data sample, and were found to be $(91.9 \pm 1.4)\%$ and $(96.2
\pm 0.3)\%$, respectively.

The systematic errors in the measurement of the branching ratios
are listed in Table~\ref{tab:syst}.
They were studied by varying the ranges of the kinematic parameters 
in the modeling of the $\Sigma^+$ production at the target, 
the assumed target positions, the magnetic fields,
and the parameters describing the $\spmumu$ and $\pi^0_{ee\gamma}$ 
decays in the MC simulations. 
The systematic errors were dominated by uncertainties 
in the modeling of the $\Sigma^+$ production, which was mainly due 
to the small normalization sample.
The systematic error includes the statistical error from the
normalization mode.

The branching ratio for the $\spmumu$ decay hypothesis is 
$[8.6^{+6.6}_{-5.4}({\rm stat}) \pm 5.5({\rm syst})] \times 10^{-8}$,
in apparent disagreement with the expected
branching ratio using the form-factor parameters given above, 
$(1.1 \pm 0.3) \times 10^{-8}$, where the error 
includes only the uncertainty in the parameter $(b_2/b_1)$. 
The branching ratio for the $\sppmumu$ hypothesis is 
$\mathcal{B}(\Sigma^{+} \ra p P^0, P^0 \ra \mu^+ \mu^-) = 
[3.1^{+2.4}_{-1.9}({\rm stat}) \pm 1.5({\rm syst})] \times 10^{-8}$.
The statistical errors for the branching ratio measurements were
estimated by using the statistical table in Ref.\ \cite{feldman}.
If the three signal events are assumed to be background events from some 
unknown source, then using the method in Ref.\ \cite{cousins} 
\begin{table}[htpb]
\centering
\caption{Fractional systematic errors ($\sigma_{B}/\mathcal{B}$) 
in the branching ratios of $\spmumu$ and of $\sppmumu$.}
\label{tab:syst}
\begin{ruledtabular}
\begin{tabular}{lrr}
   & \multicolumn{2}{c}{$\sigma_{B}/B$ (\%) \rule[-1.5mm]{0mm}{5mm}} \\
\multicolumn{1}{c}{\raisebox{1.5ex}[0pt]{Source}}
   & \multicolumn{1}{c}{~$\spmumu$~~} 
   & \multicolumn{1}{c}{~$\sppmumu$~~} \\
\hline
Normalization                                 &  14.7   & 14.7 \\
Modeling of $\Sigma^+$ production             &  54.3   & 44.6 \\
Beam targeting                                &  11.1   &  8.7  \\
Magnetic field                                &   2.2   &  3.9  \\
Trigger efficiency                            &   1.5   &  1.5  \\
Muon identification                           &   0.3   &  0.3  \\
$\spmumu$ form factor                         &  28.9   & \multicolumn{1}{c}{}   \\
$\pi^0$ form factor                           &   1.8   &  1.8  \\
$B(\sigp_ppi0)$                               &   0.6   &  0.6  \\
$B(\pi^0 \ra e^+ e^- \gamma)$                 &   2.7   &  2.7  \\
MC statistics                                 &   1.3   &  1.3  \\ \hline
Total                                         &  64.4   & 48.1 \\ 
\end{tabular}
\end{ruledtabular}
\end{table}
to take the systematic error into account,
we obtain an upper limit 
at 90\% C.L. of $\mathcal{B}(\spmumu) < 3.4 \times 10^{-7}$.

In summary, we observe three clean $\Sigma^+ \ra p \mu^+ \mu^-$ candidates. 
This is the first evidence for this decay. 
The probability that the three events have a
dimuon mass within 1 MeV/$c^2$ of each other in the SM
is estimated to be $< 1$\%. 
The three events are
consistent with the $\Sigma^{+} \ra p P^0, P^0 \ra \mu^+ \mu^-$ 
decay hypothesis, with a $P^0$ mass of $214.3 \pm 0.5$ MeV/$c^2$.

\begin{acknowledgments}
We thank the staffs of Fermilab and the participating institutions for their
contributions. We acknowledge many useful discussions with D.S. Gorbunov and P.
Singer. This work was supported by the U.S. Department of Energy and the
National Science Council of Taiwan, R.O.C.  
~E.C.D. and K.S.N. were partially
supported by the Institute for Nuclear and Particle Physics. K.B.L. was
partially supported by the Miller Institute.
\end{acknowledgments}


\begin{thebibliography}{99}
\bibitem{bergstr}
L. Bergstr\"om, R. Safadi and P. Singer, Z. Phys. C~{\bf 37}, 281 (1988).

\bibitem{gor}
D. S. Gorbunov and V. A. Rubakov, Phys. Rev. D~{\bf 64}, 054008 (2001).

\bibitem{ang}
G. Ang {\em et al.}, ZPHY~{\bf 228}, 151 (1969).

\bibitem{corrigan}
D. Corrigan and N.N. Trofimenkoff, Nucl. Phys. B~{\bf 40}, 98 (1972).

\bibitem{burnstein}
R. A. Burnstein {\sl et al.}, hep-ex/0405034 (2004).

\bibitem{pdg}
S. Eidelman {\em et al.} (Particle Data Group),
Phys.\ Lett.\ B {\bf 592}, 1 (2004).

\bibitem{mikaelian}
K. O. Mikaelian and J. Smith, Phys. Rev. D~{\bf 5}, 1763 (1972).

\bibitem{feldman}
G. J. Feldman and R. D. Cousins, Phys. Rev. D~{\bf 57}, 3873 (1998).

\bibitem{cousins}
R. D. Cousins and V. L. Highland, Nucl. Instrum. Methods 
Phys. Res., Sec. A~{\bf 320}, 331 (1992).

\end{thebibliography}
\end{document}